\documentclass[11pt]{article}

\usepackage[iso-8859-7]{inputenc}
\usepackage{epsfig}
\usepackage{graphicx}
\usepackage{epstopdf}
\usepackage{amsfonts}
\usepackage{amssymb}
\usepackage{amsthm}
\usepackage{amsmath}
\usepackage{multirow}

\newcommand{\newc}{\newcommand}
\newc{\ra}{\rightarrow}
\newc{\lra}{\leftrightarrow}
\newc{\be}{\begin{equation}}
\newc{\ee}{\end{equation}}
\newc{\ba}{\begin{eqnarray}}
\newc{\ea}{\end{eqnarray}}
\newc{\ov}{\overline}
\newc{\pa}{\partial}
\newc{\D}{\Delta}

\newc{\nn}{\nonumber}

\begin{document}
\thispagestyle{empty}

\hfill OUTP-10-27P

\hfill CERN-PH-TH/2010-208
\vskip 2truecm
\vspace*{3cm}
\begin{center}
{\Huge {\bf
 Yukawa couplings and fermion mass structure in F-theory GUTs}}\\
\vspace*{1cm}
{\bf
G.K. Leontaris$^{1}$,
G.G. Ross$^{2,3}$}\\
\vspace{4mm}
$^1$ Physics Department, Theory Division, Ioannina University, \\
GR-45110 Ioannina, Greece\vspace{1mm}\\
$^2$ Department of Physics, CERN Theory Division, \\
CH-1211, Geneva 23, Switzerland\vspace{1mm}\\
$^3$ Rudolf Peierls Centre for Theoretical Physics,\\
University of Oxford, 1 Keble Road, Oxford, OX1 3NP, UK
\end{center}

\vspace*{1cm}
\begin{center}
{\bf Abstract}
\end{center}
\noindent
The calculation of Yukawa couplings in F-theory GUTs is developed. The method is applied to the top and bottom Yukawa couplings in an $SU(5)$ model of fermion masses based on family symmetries coming from the $SU(5)_\perp$ factor in the underlying $E(8)$ theory. The remaining Yukawa couplings involving the light quark generations are determined by the Froggatt Nielsen non-renormalisable terms generated by heavy messenger states. We extend the calculation of Yukawa couplings to include massive states and estimate the full up and down quark mass matrices in the $SU(5)$ model. We discuss the new features of the resulting structure compared to what is usually assumed for Abelian family symmetry models and show how the model can give a realistic quark mass matrix structure.  We extend the analysis to  the neutrino sector masses and mixing where we find that tri-bi-maximal mixing is readily accommodated. Finally we discuss mechanisms for splitting the degeneracy between the charged leptons and the down quarks and the doublet triplet splitting in the Higgs sector.

\section{Introduction}

There has been extensive discussion of the structure of fermion masses in
F-theory~\cite{Donagi:2008ca}-\cite{Blumenhagen:2009up}. Two  ways have been suggested to generate the masses and mixing angles of the remaining two generations.
In the first case all three families with the same Standard Model representation content belong to a single matter curve~\cite{Heckman:2008qa}-\cite{Conlon:2009qq}. By imposing a monodromy, rank-one mass matrices can be obtained giving the third generation quarks and charged leptons mass.  Corrections coming from non-trivial flux configurations generate the remaining entries of the mass  matrices and have been shown to give rise to a promising hierarchical mass structure. In this case, to ensure the initial rank-one structure, it is important that there is a single intersection of the matter and Higgs curves in the up, down quark and charged lepton sectors.

A second possibility~\cite{Dudas:2009hu,King:2010mq} is that some or all of the quark and lepton families are assigned to different curves and the additional $U(1)$ symmetries decending from the underlying $E(8)$ theory are then family symmetries that enforce a leading order rank one structure after imposing a monodromy. Even in the absence of flux the remaining entries of the mass matrices can be generated once the family symmetry is spontaneously broken by vacuum expectation values for  `familon' fields. These appear through non-renormalisable operators generated by the exchange of heavy fields,  Kaluza-Klein modes or vectorlike states, that acquire mass when the underlying GUT symmetry is broken. In this case there is no constraint on the number of intersections of the matter and Higgs fields.

A particularly interesting aspect of F-theory is that the Yukawa couplings can be determined from the local structure of the theory~\cite{Katz:1996xe}-\cite{Hayashi:2009ge}. This is because the matter and Higgs fields live on brane intersections and their wave functions are exponentially damped in the direction orthogonal to the intersections. As a result the Yukawa couplings, that are generated by the overlap of the matter and Higgs fields, are dominated by the region close to the intersection point of the matter and Higgs curves.  Of course the calculation of the normalisation of the fields that enters in the physical Yukawa couplings requires non-local information because it involves an integral along the matter curve where the wave function is undamped.  Assuming a very simple form for the normalisation it has been shown~\cite{Font:2009gq,Cecotti:2009zf} that the typical magnitude of a Yukawa coupling is in reasonable agreement with that required for the top quark.

In this paper we extend the calculation of Yukawa couplings to the case of F-theory GUTs with fermion mass structure organised by the $U(1)$ family symmetries. The calculation relies on the fact that  the charges of the matter and Higgs fields are determined by the underlying GUT because the charges determine the detailed form of the wave function. We determine the general form of the Yukawa couplings for a general charge structure. Calculation of the non-renormalisable operators determining the light quark masses and mixings require the calculation of Yukawa couplings between the light states and the massive Kaluza-Klein modes or massive vectorlike states that are integrated out when calculating the dimension 5 terms. We discuss the determination of the wave functions for massive states and use them to calculate these Yukawa couplings. Using this together with a  very symmetric choice for the massive spectrum and normalisations we illustrate the resulting structure by computing the full matrix of Yukawa couplings in a specific $SU(5)$ model that has been shown to have phenomenologically promising properties.

We consider in detail the phenomenological implications of the resulting structure. Because the Yukawa couplings are correlated there can be strong cancellations in the determinant and sub-determinants of the fermion mass matrices and this strongly affects  the possible mass structures. In addition we find that significant cancellations can occur between the up and down quark contribution to mixing angles. Both these features go beyond what is normally assumed in family symmetry models. We discuss how texture zeros following from an underlying geometrical structure can lead to viable quark mass structure.

We consider the neutrino mass structure in the illustrative model and show how it may lead to the near bi-tri-maximal mixing found in neutrino oscillation experiments.
The extension to charged leptons is problematic in models with an $SU(5)$ GUT structure due to the degeneracy between down quarks and leptons and we consider two possible resolutions of the problem. The first involves radiative threshold corrections to the down quark mass matrix that, together with a strong cancellation in a sub-determinant of the down quark mass matrix, can readily split the muon mass from the strange quark mass while preserving the ``good'' $SU(5)$ mass relations. The second involves flux splitting of $SU(5)$ representations. For the case such flux splitting splits the doublets from the triplets in the Higgs sector the resulting non-degeneracy in the massive Higgs sector breaks the quark lepton degeneracy in the non-renormalisable terms of the mass matrix. We present a simple example of this mechanism. We also consider an example in which the flux splits the matter representations  so that down quarks and leptons may live on different matter curves and we consider in detail how the method can be applied to the second generation only.

\section{F-theory structure}

In this section  to facilitate our subsequent analysis and set up our notation, we review the basic formalism following~\cite{Beasley:2008dc}~\footnote{For recent reviews in F-theory see~\cite{Denef:2008wq,Weigand:2010wm}.}.

In  10-dimensions the super Yang-Mills (YM) theory, consists of the gauge field and the adjoint-valued  fermion
in the positive-chirality spinorial ${\bf 16_+}$ representation of $SO(9,1)$. Under the reduction to $R^{(7,1)}$
these decompose to an eight-dimensional gauge field $A$, two real scalars $\Phi_{8,9}$ (usually combined
to $\varphi/\bar\varphi=\Phi_8\pm\imath\Phi_9$) and two $SO(7,1)$ fermions $\Psi_{\pm}$ conjugate to each other.
Moreover, there is an $R$-symmetry of the eight-dimensional  theory emerging under the reduction $SO(9,1)\ra SO(7,1)\times U(1)_R$.

F-theory is described~\cite{Beasley:2008dc} by an eight-dimensional YM theory on $R^{(3,1)}\times S\; (\,=R^{(7,1)})$, where $S$ is a
four dimensional K\"ahler surface  wrapped by the 7-brane and is supposed to support a unified gauge  group of ADE type denoted by $G_S$ .

We choose two complex  coordinates $z_1,z_2$ to parametrise the surface $S$ while we assume that the
canonical  form is
\ba
\omega&=&\frac{\imath}{2}\left(dz_1\wedge d\bar z_1+dz_2\wedge d\bar z_2\right)
\ea
To preserve $N=1$ sypersymmetry in D=4 the YM theory must be topologically twisted~\cite{Beasley:2008dc}
since $S$ has a non-trivial canonical bundle and spinors under local $SO(4)$ rotations are not well defined globally.
 Because we assume that the hypersurface $S$ has a K\"ahler form, the 8-dimensional YM theory admits a unique  topological
 twist on $R^{(3,1)}\times S$ -specified by the embedding of $U(1)_R$ into the invariant $U(2)$ subgroup of $SO(4)$-
 which preserves $N=1$ SUSY. Under the twisted theory, the scalars and fermions appear as
forms. In particular, the scalar $\varphi$ transforms on $S$ as a section $\Omega_S^2\otimes {\rm ad}(P)$,
where $\Omega_S^p$ stands for the holomorphic $p$-form and $P$ the principal bundle in the adjoint representation
\ba
\varphi&=&\varphi_{mn}\;dz^m\wedge dz^n
\ea
and analogously for the conjugate $\bar\varphi$. Similarly, the fermions appear as holomorphic
(or antiholomorphic) forms of type $(p,0)$ or $(0,p)$ with $p=0,1,2$. These are $\eta_{\alpha}$
transforming as a section of ${\rm ad}(P)$ (i.e., is a $(0,0)$ form), $ \bar\psi_{\dot{\alpha}} =\bar\psi_{\dot{\alpha}m}dz^m$
as section of $\Omega_S^1\otimes {\rm ad}(P)$, $\chi_{\alpha}=\chi_{\alpha mn}dz^m\wedge dz^n$ as
section of  $\Omega_S^2\otimes {\rm ad}(P)$ and analogously for their complex conjugates. At the
$d=4, N=1$ level, the fields  are organized as one gauge and two chiral  multiplets as follows
\ba
(A_{\mu},\eta),\; (A_{\bar m},\psi_{\bar m}),\; (\phi_{12},\chi_{12})
\ea
and similarly for their complex conjugates.

The structure of the D=4 effective theory theory is derived from the D=8 effective action.
The equations of motion for the zero modes are derived by taking variations of the action with
respect to $\eta,\psi,\chi$~\cite{Beasley:2008dc}. These are:
\ba
\omega\wedge\partial_A\psi+\frac{\imath}{2}[\bar\varphi,\chi]&=&0
\\
\bar\partial_A\chi-2\imath\sqrt{2}\omega\wedge\eta-[\varphi,\psi]&=&0\label{eqmot}
\\
\bar\partial_A\psi-\sqrt{2}[\bar\varphi,\eta]&=&0
\ea

Fields are located on matter curves $\Sigma$ following the intersection of surfaces $S$ and $S'$. The number of zero modes on $\Sigma$ is given by topological invariants and in this paper we assume that the quark and lepton families  belong to different curves. We use the description of such fields introduced in~\cite{Beasley:2008dc,Font:2009gq}. The D=8 theory on S has gauge group $G_{\Sigma}$ and it is broken by turning on a background for the adjoint scalar $\varphi$ with the form
\ba
\langle\varphi\rangle&=&m^2\;z\; Q
\ea
where $z$ is a complex coordinate of S and  $Q$ is a generator of the $U(1)$ in $G_{\Sigma}$ that is carried by the $S'$ brane. We have introduced mass parameters so that $\varphi$ has the standard dimension. It is expected to be of the order of the F-theory mass scale $M_*$. The locus $z_1=0$ defines the curve $\Sigma$. On it the gauge group is $G_{\Sigma}$ but this is broken to $G_S\times U(1)$ off the curve.

\section{Yukawa couplings}
\subsection{Zero mode wavefunctions}\label{ZM}
Yukawa couplings occur at the intersection of three matter curves.  To describe the interacting fields at the intersection it is necessary to introduce a more general form for the background for the adjoint scalar $\varphi$

\ba
\langle\varphi\rangle&=&m^2(z_1Q_1+z_2Q_2)
\label{adjoint}
\ea
where $Q_1,Q_2$  are the $U(1)$ generators of the enhanced gauge symmetry at the intersection.  We may always choose the basis such that one matter curve, $\Sigma_1$, corresponds to the locus $z_1=0$ and on it the group is enhanced to $G_{\Sigma_1}\supset G_S\times U(1)_1$. The states living on this curve are charged only under the first $U(1)$. The second matter curve $\Sigma_2$ corresponds to the locus  $q_1z_1+q_2z_2=0$ where $q_{1,2}$ are the charges of the states living on $\Sigma_2$ and on it  the unbroken group is $G_{\Sigma_2}\supset G_S\times U(1)_a$ where $U(1)_a=(q_1U(1)_1+q_2U(1)_2)/\sqrt{q_1^2+q_2^2}$.  The third matter curve, $\Sigma_3$ has locus $q_1'z_1+q_2'z_2=0$ where  the charges $q_1',\; q_2'$  cancel the charges on the first two matter curves.

Matter fields have wave functions peaked along $\Sigma_i$  and transform as bi-fundamentals under $G_S\times U(1)_i$. The equations of motion for the zero modes with charges $q_1,q_2$ are given by~\cite{Donagi:2008ca,Beasley:2008dc,Font:2009gq,Grimm:2010ks}
\ba
\pa_1\psi_1+\pa_2\psi_2-m^2(q_1\bar z_1+q_2 \bar z_2)\chi&=&0\label{de1}\\
\bar\pa_1\chi-m^2(q_1\, z_1+q_2 \, z_2)\psi_1&=&0\label{de2}\\
\bar\pa_2\chi-m^2(q_1\, z_1+q_2 \, z_2)\psi_2&=&0\label{de3}
\ea
These are readily rotated to the canonical form by the rotations \ba
w&=&\cos\theta\,z_1+\sin\theta\,z_2\\
u&=&-\sin\theta\,z_1+\cos\theta\,z_2
\ea
and
\ba
\psi_w&=&\sin\theta\,\psi_2+\cos\theta\,\psi_1\\
\psi_u&=&\cos\theta\,\psi_2-\sin\theta\,\psi_1
\ea
where $\tan\theta =q_2/q_1$ and $q=\sqrt{q_1^2+q_2^2}$.

Equs (\ref{de1}-\ref{de3}) become
\ba
\pa_w\psi_w+\pa_u\psi_u&=&m^2 q\,\bar w\,\chi\\
\bar\pa_{\bar w}\chi&=&m^2 q\,w\,\psi_w\\
\bar\pa_{\bar u}\chi&=&m^2q\,u\,\psi_u
\ea
For localised solutions set $\psi_u=0$, then $\bar\pa_{\bar u}\chi=0$ and the differential equations  reduce to
\ba
\pa_w\psi_w\,=\,m^2q\,\bar w\,\chi&,& \bar\pa_{\bar w}\chi\,=\,m^2q\,w\,\psi_w
\ea
These are solved taking  $\chi\propto e^{-\lambda |w|^2}$ giving  $\bar\pa_{\bar w}\chi\,=-\lambda \,w\chi =m^2qw\psi_w$, so
\ba
\psi_w&=&-\frac{\lambda}{m^2q}\chi\,=\, -\frac{\lambda}{m^2q}\, e^{-\lambda |w|^2}.
\ea
The parameter $\lambda$ is determined when $\psi$ is substituted in the first equation above
\[
\pa_w\psi_w\;=\;\frac{\lambda^2}{m^2q}\,\bar w\, e^{-\lambda |w|^2}\;=\;m^2q\bar w\,\chi
\]
which implies $\lambda=\pm m^2 q\equiv \pm m^2\sqrt{q_1^2+q_2^2}$. Choosing the positive root the wave function is
\ba
\psi_w&=&e^{-m^2q|w|^2}=exp\{-m^2 q\left|\cos\theta z_1+\sin\theta z_2\right|^2\}\nn\\
      &=&e^{-m^2\frac{|q_1z_1+q_2z_2|^2}{q}}\label{psiw}
      \label{wf}
\ea
 i.e., the wave function exhibits a Gaussian profile and thus it falls off exponentially away from the curve.

 This discussion has been concerned with the dependence of the wave function on the co-ordinates transverse to the matter curve. There is also an unknown dependence on the longitudinal coordinate $z_{L}$ so $\psi=\psi_{w}f(z_{L})$ where $f(z_{L})$ is a holomorphic function. In what follows we do not include the effect of $f(z_{L})$ on the Yukawa couplings. This is the case if $f(z_{L})$ is constant or if the matter curve intersections occur at a single point (of enhanced $E(8)$ symmetry) so that only the value of $f(z_{L})$ at this point is relevant. In both these cases the effect of $f(z_{L})$ can be absorbed in the wave-function normalisation factors. In Section \ref{CP} we comment on the changes that occur if this is not the case.

\subsubsection{Normalisation of the wave-functions}
Since the wave-functions are only localised in the direction transverse to the matter curves the normalisation of the fields is inherently a non-local property.

Consider first the normalization of a wavefunction $\psi \propto e^{-m^2q|z|^2}$. The normalisation is given by $\frac{1}{\sqrt{\cal C}}$
\ba
{\cal C}&=&M_*^4\int_S\,|\psi|^2dz\wedge d\bar z .
\ea
This gives
\ba
{\cal C}=\pi \frac{M_*^4}{m^2q}R^2
\ea
where the factor $\pi \frac{M_*^4}{m^2q}$ comes from the (local) integration over transverse coordinates and the factor $R^2$ parameterises the (non-local) integral over the longitudinal coordinates.  We expect that $m=O(M_*)$ and, for the case the wave function has no structure along the longtitudinal direction, we have $r=R$ where $R$ is the curvature of the hypersurface $S$ with $M_*^2R^2\sim\alpha_G$. In our estimates of Yukawa couplings presented below we will use this approximation and take the normalisation of the matter fields to be

 \ba
\frac{1}{\sqrt{\cal C}}&=&\left(\frac{q}{\pi}\right)^{\frac 12}\frac{1}{M_*R}=
\left(\frac{q}{\pi}\right)^{\frac 12}a_G^{\frac 14}.
\label{norm}
\ea
It should be stressed however that this form assumes a very symmetric structure of the compactification manifold and wave functions

\subsection{Trilinear couplings}
We are interested in determining the Yukawa couplings involving two matter fields either quarks or leptons, and a Higgs field. In F-theory  these couplings are computed in terms of overlap
integrals~\cite{Strominger:1985it}
\ba
\lambda_{ij}&=&\frac{M_*^4}{(2\pi)^2}\int_S\psi_i\psi_j\phi\; dz_1\wedge\,d\bar z_1\wedge dz_2\wedge\,d\bar z_2
\ea
where $\psi_i,\phi$ are the internal  wave-functions of the relevant fermions and Higgs field
of a particular coupling. As we have discussed these wave-functions are strongly peaked along the matter curve and so one can get an accurate estimate of the integral using only the local form of the wave-functions close to the intersection point. For the calculations in specific models it is useful to determine the overlap integrals in an arbitrary basis. In this case, using eq(\ref{wf}),  the general form of the integral is
\ba
I=\frac{M_*^4}{(2\pi)^2}\int\,e^{-m^2 |xz_1+yz_2|^2}\,e^{-m^2|uz_1+vz_2|^2}\,
e^{-m^2|az_1+bz_2|^2}\,d^2z_1\,d^2z_2
\ea
where
\ba
x=\frac{q_1}{\sqrt{q}},\;y=\frac{q_2}{\sqrt{q}},\;u=\frac{q'_1}{\sqrt{q'}},\;v=\frac{q'_2}{\sqrt{q'}},\;
a=\frac{q''_1}{\sqrt{q''}},\;b=\frac{q''_2}{\sqrt{q''}}\\
q=\sqrt{{q_1}^2+{q_2}^2},\; q'=\sqrt{{q'_1}^2+{q'_2}^2}, \;q''=\sqrt{{q''_1}^2+{q''_2}^2}
\ea
By charge conservation, $q''_i=-q_i-q'_i,\;i=1,2$, giving
\[
a=-\frac{q_1+q_1'}{\sqrt{q''}},\;b=-\frac{q_2+q_2'}{\sqrt{q''}},
\]

The integral is readily performed giving
\ba
I&=&\frac{M_*^4}{m^4}\frac{qq'q''}{q+q'+q''}\,\frac{1}{(q_1q_2'-q_1'q_2)^2}\label{res}
\ea
As before we expect $M_*\sim m$ so, setting them equal, and using the normalisation of eq(\ref{norm}) the Yukawa coupling is
\ba
\lambda&=&\frac{(2\pi)^2}{\pi^{3/2}}\,(qq'q'')^{1/2}a_G^{3/4}\,I\;=\;\frac{4\,\sqrt{\pi\,a_G^{3/2}} }{q+q'+q''}\,\frac{(qq'q'')^{3/2}}{(q_1q_2'-q_1'q_2)^2}
\label{Yukint}
\ea

 \subsection{Yukawa couplings involving massive modes modes}\label{massiveyuks}

In the previous Section we determined the Yukawa couplings for zero modes. However, as discussed below, we are also interested in the coupling of massless to massive modes as these are involved in the generation of higher dimension operators and the latter are important in determining the phenomenology of the effective low energy theory.  The massive modes of interest may be vectorlike states that gain mass at a stage of spontaneous breakdown or they may be Kaluza Klein (KK) excitations.

 In the first case the mass will be determined by the scale of symmetry breaking. In the case of KK modes, in general the explicit mass formula of the KK-modes is unknown. In a Calabi-Yau manifold, they
 can be parametrised as $m_{KK}^{_2}\sim g_K(t,S_1){\cal E}_K$ where  $S_1$ is the real part of
 the dilaton field $S=e^{-\phi}+\imath \,C_0$, $t$ is the 2-cycle volume, while the function $g(t,S_1)$
  determines the scale and ${\cal E}_K$ is an unknown function of the moduli.
  For the two-cycle $t_i$\footnote{$t_i$ is connected to the overall volume  $V=t_i\tau_i$, where
   $\tau_i$ is the corresponding 4-cycle related to the K\"ahler moduli $T_i=\tau_i+\imath b_i$}
   transverse to the 7-brane we may assume $g\sim t_i$. In the presence of fluxes ${\cal A}$, the mass formula receives corrections of the form~\cite{Berg:2007wt}
 \[
 m_{KK}^2\sim \frac{1}{{\cal E}^K}\frac{1}{t_{str}\left(1+\frac{{\cal A}^2}{t_{str}^2}\right)}
 \sim  \frac{1}{{\cal E}^K} \frac{\sqrt{S_1}}{t}\left(1-\frac{{\cal A}^2}{t^2}S_1\right)
 \]
In what follows we shall assume constant fluxes and that the massive modes have a mass smaller that the compactification scale determined by the specific curvature of the hypersurface $S$.

It is straightforward to determine the wavefunction of a  massive state,  $\psi_n$, for the case that the mass, $M$, is generated
by the vev of some internal gauge field $A_n$. The relevant equation is modified, so that
\ba
 (\partial_1-\imath A_1) \psi_1-m^2q_1\bar z_1\chi&=&0\nn
 \\
 \bar\partial_{\bar 1}\chi -m^2q_1\, z_1\psi_1&=&0\nn
\ea
$\psi_1$ admits a solution
\[\psi_1\sim {\rm exp}\left[\left(M^2-\sqrt{M^4+q_1^2m^4}\right)|z_1|^2\right]\sim
 {\rm exp}\left[\left(-q_1m^2+M^2\left(1-\frac{M^2}{2q_1m^2}\right)|z_1|^2\right)\right]
\]
Defining
\ba
\rho =\frac{M^2}{q_1m^2}<1,\;\;\xi\approx 1-\rho+\frac{1}{2}\rho^2,
\label{xi}
\ea
the wave function has the form $\psi_1\sim e^{-q_1m^2\xi|z_1|^2}, \;\;\xi<1$.

\subsubsection{The coupling involving massive modes}

It is now straightforward to determine the Yukawa coupling involving this state and two zero modes by
replacing the mass parameter $m^2$ of one exponent in the trilinear coupling by $m^2\xi$. This gives
\ba
\lambda&=&\frac{(2\pi)^2}{\pi^{3/2}}\,(qq'q''\xi)^{1/2}a_G^{3/4}\,I\;=\;4\,(\pi qq'q''\xi)^{1/2}a_G^{3/4}\,I_{\xi}\nn
\\
I_{\xi}&=&\frac{qq'q''}{(q+q')\xi+q''}\,\frac{1}{(q_1q_2'-q_1'q_2)^2}\label{resxi}
\ea
and so
\ba
\lambda(\xi)&=&\frac{4\sqrt{\pi\xi a_G^{3/2}}}{(q+q')\xi+q''}\frac{(qq'q'')^{3/2}}{(q_1q_2'-q_1'q_2)^2}.
\label{Yukawa}
 \ea
To generalise for more than one KK-modes with different masses we introduce
$\xi,\xi',\xi''$ parameters in (27) as follows
\ba
x=\sqrt{\xi}\frac{q_1}{\sqrt{q}},\;y=\sqrt{\xi}\frac{q_2}{\sqrt{q}},\;u=\sqrt{\xi'}\frac{q'_1}{\sqrt{q'}},\;v=\sqrt{\xi'}\frac{q'_2}{\sqrt{q'}},\;
a=\sqrt{\xi''}\frac{q''_1}{\sqrt{q''}},\;b=\sqrt{\xi''}\frac{q''_2}{\sqrt{q''}}
\ea
This is essentially equivalent to the replacements
\[
q\ra \frac{q}{\xi}, q'\ra\frac{q'}{\xi'},q''\ra\frac{q''}{\xi''}\]
so that the generalization of our integral (32) is
\[
I_{\xi}=\frac{qq'q'' }{\xi'\xi'' q+\xi\xi''q'+\xi\xi'q''}\,\frac{1}{(q_1q_2'-q_1'q_2)^2}.
\]
For two KK-modes and one zero mode we have (setting $\xi''=1$)
\[
I_{\xi}=\frac{qq'q'' }{\xi' q+\xi q'+\xi\xi'q''}\,\frac{1}{(q_1q_2'-q_1'q_2)^2}.
\]

\section{ Quark and lepton masses and mixing angles in a semi-realistic $SU(5)\times U(1)^3$ model}

\subsection{The model}

In \cite{Dudas:2009hu} and \cite{King:2010mq} semi-realistic F-theory GUT models, capable of generating reasonable structure for quark and lepton masses were constructed. The models assign families to different matter curves and as a result the Abelian factors left after imposing the monodromy necessary to get near-rank-one structure for quarks and charged leptons become family symmetries. They are spontaneously broken by familon vevs and are able to control the hierarchy even for the case the Yukawa couplings arise from  multiple intersections.  Here we construct a variant of the model constructed in \cite{Dudas:2009hu}  that, in addition to realistic quark and charged lepton mass matrices, can give near tri-bi maximal mixing in the neutrino sector via the see-saw mechanism.

\begin{table}[h] \centering%
\begin{tabular}{|c|c|c|c|}
\hline
\textbf{Field} & $SU(5)\times SU(5)_{\perp}\;$\textbf{Representation} & $\mathbf{SU(5)}_{\perp }$ component &
R-parity\\ \hline
$Q_{3},U_{3}^c,l^c_3$ & $\left( 10,5\right) $ & $t_{1,2}$&$ -$ \\ \hline
$Q_{2},U_{2}^{c},l^c_2$ & $\left( 10,5\right) $ & $t_{4}$& $-$ \\ \hline
$Q_{1},U_{1}^{c},l^c_1$ & $\left( 10,5\right) $ & $t_{3}$& $-$ \\ \hline
$D_{3}^{c},L_3$ & $(\overline{5},10)$ & $t_{1,2}+t_{4}$&$-$ \\ \hline
$D_{2}^{c},L_2$ & $(\overline{5},10)$ & $t_{1,2}+t_{3}$ &$- $\\ \hline
$D_{1}^{c},L_1$ & $(\overline{5},10)$ & $t_{3}+t_{4}$& $- $\\ \hline
$H_{u}$ & $\left( 5,\overline{10}\right) $ & $-t_1-t_2$&$ +$ \\ \hline
$H_{d}$ & $\left( \overline{5},10\right) $ & $t_{3}+t_{5}$&$+$\\ \hline
$\theta _{ij}$ & $\left( 1,24\right) $ & $t_{i}-t_{j}$ &$+$\\ \hline
$\theta'_{ij}$ & $\left( 1,24\right) $ & $t_{i}-t_{j}$ &$-$\\ \hline
\end{tabular}%
\caption{Field representation content under $SU(5)\times SU(5)_{\perp}$} and R-parity
\label{Reps}
\end{table}

The starting point is  the $SU(5)\times SU(5)_{\perp }\subset E_8$ group. We label the weights of
$SU(5)_\perp$  by $t_i,\;i=1,\dots, 5$. To get rank 1 matrices in the family symmetry limit we impose a  monodromy
group $Z_{2}$ relating $t_{1}\leftrightarrow t_{2}$.  The matter fields tranform as bi-fundamentals under the gauge group associated with the intersecting branes generating the matter curve. To avoid $D=4$ baryon and lepton number violating terms~\cite{Hayashi:2009bt,Hayashi:2010zp,King:2010mq,Grimm:2010ez,Kuflik:2010dg}  it is necessary to include an R-parity. In~\cite{Hayashi:2009bt} it was shown that this can arise if the Calabi-Yau manifold has a $Z_2$ symmetry provided the flux also respects the symmetry. We assign the quarks and Higgs fields to the curves as shown in Table \ref{Reps}.  We assign the quarks and Higgs fields to the
curves as shown in Table \ref{Reps}. Due to the monodromy the fields with component labeled $t_{1,2}$ can either be represented by $t_1$ or $t_2$. R parity even $SU(5)$ singlets $\theta_{ij}$
belonging to the $(1,24)$ representation will play the role of familon fields while R parity odd singlets $\theta_{ij}'$ have the appropriate quantum numbers to provide right-handed neutrino states. D=5 operators that can generate nucleon decay have the form $10.10.10.\bar{5}$ are forbidden because, {\it c.f.} Table \ref{Reps}, they require a field transforming as $t_{5}$.

\subsection{The $\mu$-term and doublet triplet splitting}\label{mu}

In the MSSM a $\mu$ term, $\mu H_u H_d$, in the superpotential is necessary to have  to give mass to the Higgsinos. However, in the symmetry limit, the $\mu$ term is forbidden because there is no familon carrying $t_5$ component. There are two ways that a $\mu$ term can be generated when the symmetry is broken spontaneously.

The first is through a D-term coupling $H_uH_d\theta_{14}\theta_{43}\theta_{51}^{\dagger}$. Here $\theta_{14,43}$ are familon fields needed to generate an acceptable quark mass matrix (see next Section).  The field $\theta_{51}$ should get an F-term breaking supersymmetry and generating the $\mu$ term via the Giudice-Masiero mechanism \cite{Giudice:1988yz}. However its A-component should have a zero or very small vev to avoid introducing large nucleon decay terms.

The second possibility is that a $\mu$ term is generated via the term $\lambda H_uH_d\theta _{14}\theta_{43}\theta_{15}$. For the case $\theta_{15}$ acquires a large vev it is necessary for the coupling $\lambda$ to be very small to get an acceptably small $\mu$ term. This can happen if one or more of the couplings responsible for the higher dimension involve fields which do not intersect because then there is an exponential suppression of the coupling due to the fall-off of the wavefunction in a direction transverse to the matter curve~\cite{Katz:1996xe}.
It has been argued in ~\cite{Tatar:2009jk} that the suppression of a vertex due to this effect is $\le O(10^{-7})$ that by itself would not be sufficient. However for the case of a higher dimension operator, such as the one here, several vertices are involved so two or more may be geometrically suppressed allowing for a suppression sufficient to give a $\mu$ term of electroweak breaking order as is necessary.

An associated problem in a GUT model is the need to split the Higgs multiplets $H_{u}$ and $H_{d}$. These transform as $5$ and $\bar{5}$ representations under $SU(5)$ and contain colour triplets as well as the electroweak Higgs doublets needed to spontaneously break the electoweak symmetry. If light, the colour triplets spoil gauge coupling unification. In addition they can mediate rapid proton decay through dimension 5 operators. For these reasons they must be eliminated or given a large mass while leaving the doublet components light. There are two mechanisms to do this. One involves hypercharge flux that can split multiplets; an example of this will be given in Section \ref{flux}. The second possibility, when the compactification involves a Calabi Yau manifold,  is through discrete Wilson line breaking. Unlike flux breaking, it projects out the colour triplet zero mode states and so has no massive states to spoil gauge unification.

\subsection{Quark mass matrices}
After identifying the monodromy group the residual gauge group structure is $SU(5)\times U(1)^3$. The structure of the mass matrices depends on the choice of familon fields which acquire vevs to spontaneously break the symmetry.  Identifying $\theta_{14}$ and $\theta_{43}$ as two familons\footnote{It is shown in the Appendix that this choice is consistent with F- and D- flatness.} the quark mass matrices have the form~\cite{Dudas:2009hu}
\ba
M_{d}&=&
\left(\begin{array}{ccc}
\theta_{14}^2\theta_{43}^2&\theta_{14}\theta_{43}^2&\theta_{14}\theta_{43}\\
\theta_{14}^2\theta_{43}&\theta_{14}\theta_{43}&\theta_{14}\\
\theta_{14}\theta_{43}&\theta_{43}&1
\end{array}
\right)v_{b},
\;\;\;
M^{u}=
\left(\begin{array}{ccc}
\theta_{14}^2\theta_{43}^2&\theta_{14}^2\theta_{43}&\theta_{14}\theta_{43}\\
\theta_{14}^2\theta_{43}&\theta_{14}^2&\theta_{14}\\
\theta_{14}\theta_{43}&\theta_{14}&1
\end{array}
\right)\,v_u
\label{massmatrices}
\ea
where $v_{d,u}$ are the down and up Higgs vevs and $\theta_{ij}$ refers to the familon vevs.  Here we have suppressed the couplings of $O(1)$; they will be estimated in the next Section. We have also suppressed the masses of the messenger fields associated with the higher dimension terms.

\subsubsection{Renormalisable couplings}
In the model under consideration, in the family symmetry limit,  both up and down quark mass matrices are rank one in good agreement with the observed hierarchical mass matrix structure. To determine these masses it is necessary to determine  the top and bottom Yukawa couplings corresponding
to the coefficients of the $\{33\}$ entries of $M^d$ and $M^u$.
\begin{figure}[h]
\centering
\includegraphics[scale=0.7]{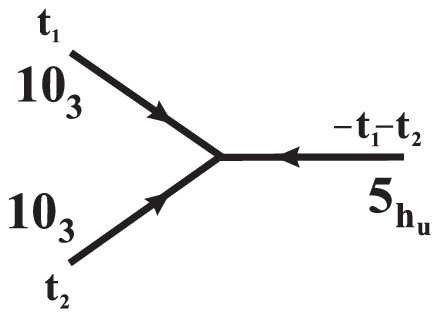}\;\;\;\;\;\;
\includegraphics[scale=0.7]{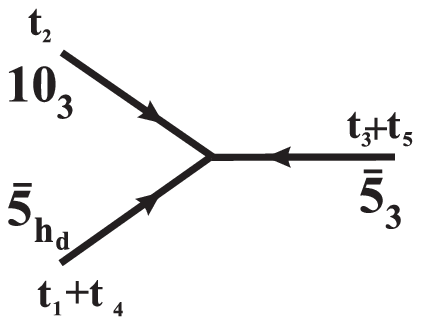}
\caption{Tree level  graphs for the top and bottom Yukawa couplings $\lambda^{t,b}_{33}$.}
 \label{top}
\end{figure}

We start with the top Yukawa coupling. The relevant graph is depicted in figure (\ref{top}) where under
the monodromy $t_1\leftrightarrow t_2$  the corresponding two tenplets $10_{t_1}, 10_{t_2}$ are identified with the third generation $10_3$. The properties of all SM fields are presented in Table \ref{Reps}. For the fields involved in a given triple intersection corresponding to a particular Yukawa coupling one combination of the charge operators has zero eigenvalue for the three fields involved. We proceed by identifying the $U(1)$ charges of the fields involved in the intersection.

The  generators of the $U(1)$s are given by the  four diagonal generators, $Q_i$,  of $SU(5)_{\perp}$. It is convenient to introduce the basis column vectors ${|t_i>}_j=\delta_{ij}$, $i,j=1,\dots 5$. The charges of a given field appearing in Table \ref{Reps} are then given by acting with  $Q_i$ on the combination of  $|t_i>$ that appear in the Table.

For a given Yukawa coupling the fields involved will have zero eigenvalue for two combinations of the charge generators. We proceed by identifying them and then forming the orthogonal charge operators $Q_1,\;Q_2$. The charges of these fields under these operators are then readily found and used in computing the Yukawa couplings from eq(\ref{Yukawa}).

Consider first the calculation of the bottom quark Yukawa coupling. The interacting fields have weights $t_{1,2},\;t_{1,2}+t_4$ and $t_3+t_5$. We first calculate the Yukawa couplings in the covering theory and then project them in the quotient space.  In the covering theory the bottom quark Yukawa couplings correspond to the couplings of fields with weights $t_1,\;t_2+t_4,\;t_3+t_5$ and $t_2,\;t_1+t_4,\;t_3+t_5$.  To calculate the coupling involving the first set we choose the orthogonal basis for the $Q_i$ given by
\ba
Q_1&=&\frac{1}{\sqrt{6}}\{2,-1,0,-1,0\}\nn\\
Q_2&=&\frac{1}{\sqrt{30}}\{2,2,-3,2,-3\}\nn\\
Q_3&=&\frac{1}{\sqrt{2}}\{0,0,1,0,-1\}\nn\\
Q_4&=&\frac{1}{\sqrt{2}}\{0,1, 0,-1,0\}
\ea

The states are annihilated by $Q_3$ and $Q_4$ so the general form of the background for the adjoint scalar is given by eq (\ref{adjoint}) with charges
\ba
\{q_1,q_2\}=\left\{\sqrt{\frac{2}{3}},\sqrt{\frac{2}{15}}\right\},
\;\{q_1',q_2'\}=\left\{-\sqrt{\frac{2}{3}},2\sqrt{\frac{2}{15}}\right\}
\label{Qdown}
\ea
The Yukawa coupling is calculated by substituting eq(\ref{Qdown}) in eq(\ref{Yukint}). This gives
\ba
\lambda&\approx&0.29.
\ea
One may readily check that the Yukawa coupling associated with the second weight structure is the same.

In the quotient theory the bottom Yukawa coupling is given by two terms with weight structure  given by $t_1/\sqrt{2},\;(t_2+t_4)/\sqrt{2},\;t_3+t_5$  and  $t_2/\sqrt{2},\;(t_1+t_4)/\sqrt{2},\;t_3+t_5$ and so the bottom Yukawa coupling is given by the mean of the two couplings in the covering theory
\ba
\lambda_{33}^b=\lambda&\approx&0.29.
\ea

For the case of the  top quark Yukawa coupling the relevant coupling in the covering theory is $t_1, \;t_2$ and $-t_1-t_2$. The appropriate basis is given by
\ba
Q_1&=&\frac{1}{\sqrt{30}}\{3,3,-2,-2,-2\}\nn\\
Q_2&=&\frac{1}{\sqrt{2}}\{1,-1, 0,0,0\}\nn\\
Q_3&=&\frac{1}{\sqrt{2}}\{0,0,1,0,-1\}\nn\\
Q_4&=&\frac{1}{2}\{0,0,1,-2,1\}
\label{Qbasis}
\ea
Again the states are annihilated by $Q_3$ and $Q_4$ so the general form of the background for the adjoint scalar is given by eq (\ref{adjoint}) with charges

\ba
\{q_1,q_2\}=\left\{\sqrt{\frac{3}{10}},\frac{1}{\sqrt{2}}\right\},
\;\{q_1',q_2'\}=\left\{\sqrt{\frac{3}{10}},-\frac{1}{\sqrt{2}}\right\}
\label{Qtop}
\ea
The Yukawa coupling is calculated by substituting (\ref{Qtop}) in eq(\ref{Yukint}), giving
\[
\lambda'\approx 0.31
\]
In the quotient theory the top Yukawa coupling is given by fields with weight structure $(t_{1}+t_{2})/\sqrt{2}$, $(t_{1}+t_{2})/\sqrt{2}$ and $ -t_1-t_2$ so
\ba
\lambda_{33}^u=\lambda'&\approx&0.31.
\ea

One sees that the magnitude of the top and bottom couplings are quite similar corresponding to the large $\tan\beta$ regime, where $\tan\beta=\frac{v_{u}}{v_{b}}$. This result is sensitive to the normalisations of the fields which involve non-local information and so is sensitive to our assumption that the effective radii of the up and down matter curves are the same.  If the couplings are indeed similar there will be significant  $\tan\beta$ enhanced threshold corrections to the down quark mass matrix entries \cite{Pierce:1996zz} that can affect the relations between the down quarks and charged leptons \cite{Ross:2007az}. Indeed, as discussed below, these could explain the difference between the muon and strange quark masses at the unification scale.

\subsubsection{ Higher dimension couplings}
As may be seen from eq(\ref{massmatrices}) the remaining terms in the quark mass matrices come from higher dimension operators involving the familon fields. These terms are generated by the exchange of massive messenger states that may either be vector-like states in the zero mode sector that acquire mass through spontaneous symmetry breaking or Kaluza-Klein states. In this section we will use the results of Section \ref{massiveyuks} to determine these non-renormalisable terms. Of course a full calculation would require detailed knowledge of the massive spectra and this is beyond the scope of this paper. Here we make the (over)simplifying assumption that the multiplet and mass structure of massive modes is the same for all matter curves. This may be expected to be the case for the KK spectra for the homogeneous compactification assumed above when determining the field renormalisations. In any case it serves to illustrate the nature of the contributions that generate the non-renormalisable operators.

\subsubsection{Calculation of the down-quark Yukawa matrix}
\begin{figure}[t]
\centering
\includegraphics[scale=0.6]{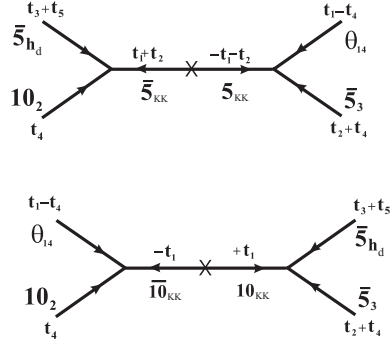}
\caption{The two graphs generating the $\lambda^{d}_{23}$ down quark Yukawa coupling.}
 \label{down13}
\end{figure}

We start with a particular example, the $\lambda^{d}_{23}$ down quark Yukawa coupling. This can be generated in the covering theory by the graphs shown in Fig. (\ref{down13}).

Consider the first vertex of the first graph and the  two orthogonal charge operators
\ba
Q_3=\frac{1}{\sqrt{6}}\{0,2,-1,0,-1\},\;
Q_4=\frac{1}{\sqrt{2}}\{0,0, 1,0,-1\}\nn
\ea
These have zero eigenvalues when acting on any of the states in the first vertex. The remaining two operators may be chosen as
\[
Q_1=\frac{1}{\sqrt{2}}\{1,0,0,-1,0\},\;
Q_2=\frac{1}{\sqrt{30}}\{3,-2,-2,3,-2\}\nn
\]

The first vertex of figure \ref{down13} is the $SU(5)$ $\ov{10}.10.1$ coupling. The external fields have weights  $t_1-t_4,\;t_4$ with the associated charge structure
\ba
\{q_1,q_2\}=\left\{-\frac{1}{\sqrt{2}},\sqrt{\frac{3}{10}}\right\},\;\{q_1',q_2'\}=\left\{\sqrt{2},0\right\}
\ea
This gives the Yukawa coupling
$\lambda_1=\frac{1.31\sqrt{\xi}}{0.89+2.31\xi}$.  For the case $\xi=1$, $\lambda_{1}=0.409$.
Repeating this it is straightforward to compute all the allowed $SU(5)$ couplings. These are tabulated in Table \ref{Yu} for the case $\xi=1$.

\begin{figure}[t]
\centering
\includegraphics[scale=0.6]{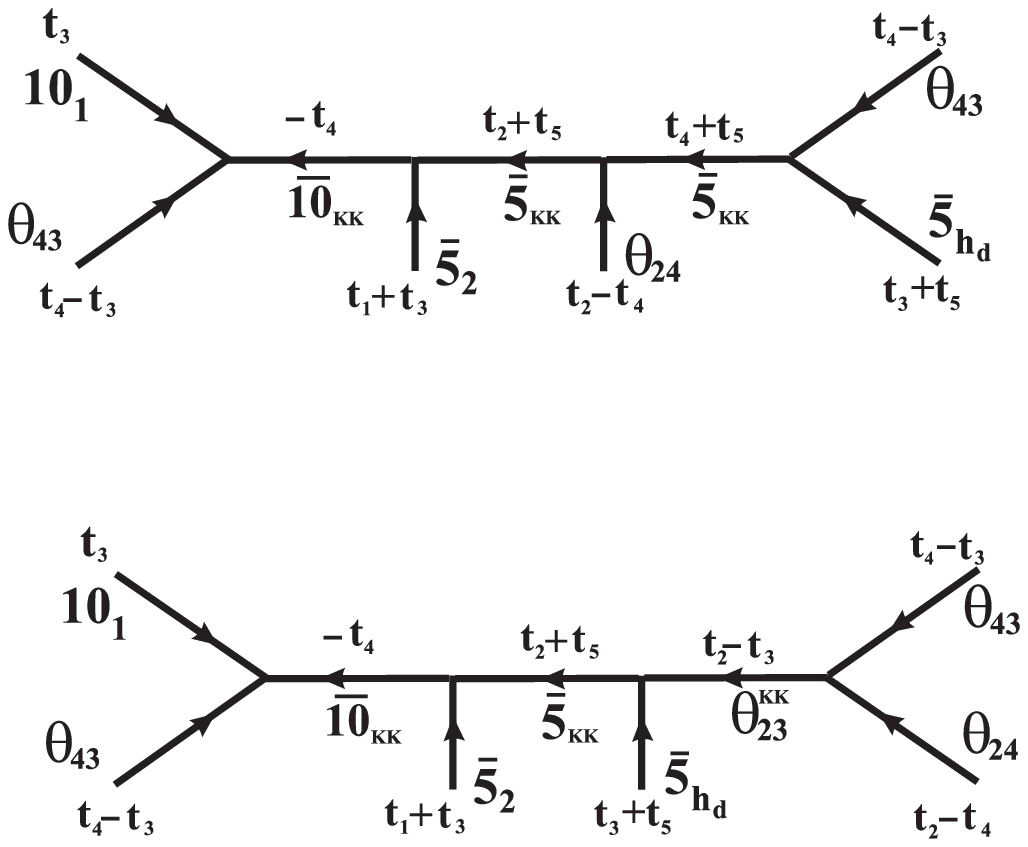}
\caption{Graphs generating the $\lambda^{d}_{12}$ down Yukawa coupling. Mass insertions in KK-mode lines are not shown to avoid clutter.}
 \label{down12}
\end{figure}

\begin{table}[h] \centering%
\begin{tabular}{|c|c|c|}
\hline
$\mathbf{SU(5)}$ \textbf{structure} & $\mathbf{U(1)}^4_{\perp }$ label &
Yukawa\\ \hline
 $ 10\cdot 10\cdot 5 $ & $t_{i},t_j,-t_{i}-t_j$&$ 0.31$ \\ \hline
 $ 10\cdot\bar 5\cdot\bar 5$ & $t_{i},t_j+t_k,t_m+t_n$& $0.295$ \\ \hline
 $\overline{10}\cdot 10\cdot 1 $ & $-t_i,t_{j},t_{j}-t_i$&$0.409$\\ \hline
  $\overline{5}\cdot 5\cdot 1  $ & $t_{i}+t_j,-t_j-t_{k},t_k-t_i$ &$0.286$\\ \hline
 $1\cdot 1\cdot 1$ & $t_{i}-t_{j},t_j-t_k,t_k-t_i$ &$0.244$\\ \hline
\end{tabular}%
\caption{The Yukawa integrals for triple intersections and their $SU(5)\times SU(1)^4_{\perp}$ structure for the case $\xi=1$.}
\label{Yu}
\end{table}

Using this one readily computes the graphs of Fig.(\ref{down13}) giving $\lambda_{23}=0.295(0.409+0.286)\theta_{14}v_{b}=0.205\theta_{14}v_{b}$ for the case $\xi=1$. In this we have suppressed the mediator mass scale, $M$, {\it i.e.} $\theta_{14}\equiv\frac{\theta_{14}}{M}$.

The remaining entries of the matrix may be computed in a similar manner. Representative graphs are shown in Figs \ref{down13} and \ref{down12}. The complete mass matrix for $\xi=1$ is given by

\ba
M_d&=&\left(
\begin{array}{lll}
0.12 \,\theta _{14}^2 \theta _{43}^2 & 0.11\, \theta _{14} \theta _{43}^2 & 0.18\, \theta _{14} \,\theta _{43} \\
0.14\, \theta _{14}^2 \theta _{43} & 0.16 \,\theta _{14} \theta _{43} & 0.20 \,\theta _{14} \\
0.09\, \theta _{14} \theta _{43} & 0.17\, \theta _{43} & 0.29
\end{array}
\right)
\label{MDD}
\ea

 Of course this result is very sensitive to our assumption that the multiplet and mass structure of massive modes is the same for all matter curves but it is nonetheless interesting to consider the implications of such a form. The results for individual matrix elements are also sensitive to the non-local structure of the compactification manifold through the normalisation factors. However the determinant and sub-determinants of the matrix are sensitive to these normalisations only through an overall factor and it is interesting to ask whether there is a substantial cancellation in these terms that is independent of the normalisation and that will significantly affect the eigenvalues of the matrix.

 \subsubsection{Phenomenological implications}\label{4.3.1}

 From eq.(\ref{MDD}) we have Det$(M_{d})=3\times 10^{-4}(\theta_{14}\theta_{43})^{2}$ and Det$(M_{d}^{23})=10^{-2}\theta_{14}\theta_{43}$ where the latter refers to the $2\times 2$ submatrix for the second and third generations that determines the product $m_{s}m_{b}$. For comparison if we impose the texture zeros in the $(1,1)$ and $(2,2)$ positions, that would follow if the matter curves determining these elements do not intersect, we find Det$(M_{d})=3\times 10^{-3}(\theta_{14}\theta_{43})^{2}$ and Det$(M_{d}^{23})=3.6\times 10^{-2}\theta_{14}\theta_{43}$. One sees that there are cancellations in the determinants of eq(\ref{MDD}). This has a significant implication for the mass matrix structure because the hierarchy of masses will not just be determined by the expansion in powers of the familon fields but also by the cancellation in the determinant and sub-determinant. As a result the familon vevs need to be chosen much larger than the case with no cancellation.  Since the model was built assuming no such cancellation one might expect the resulting structure will have problems; in particular in the original model the strong hierarchy in the up quark masses relies on small familon vevs. To avoid this we modify the model slightly by assuming texture zeros in the $(1,1)$ and $(2,2)$ positions thus avoiding the troublesome cancellations. With this and the choice $\theta_{14}=0.07$ and $\theta_{43}=0.6$  we have $m_{s}/m_{b}=0.36 \;\theta_{14}\theta_{43}=0.07$,  $m_{d}/m_{b}=3\times 10^{-3} (\theta_{14}\theta_{43})^{3}=5\times 10^{-4}$ and $V^{d}_{{cb}}\approx 0.05$ in good agreement with the measured values continued to the unification scale. However the down quark contribution to $V^{d}_{us}$ is large $V_{us}^{d}=0.5$ so that there must be a significant cancellation with the contribution from the up quark sector. To address this possibility we must determine the structure of the up quark mass sector.

\subsubsection{Calculation of the up-quark Yukawa matrix}
\begin{figure}[t]
\centering
\includegraphics[scale=0.6]{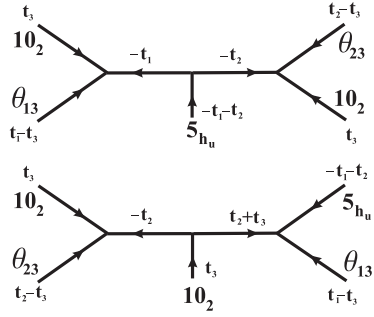}
\caption{Graphs generating the  $\lambda^{u}_{22}$ up Yukawa coupling.}
 \label{up22}
\end{figure}

\begin{figure}[t]
\centering
\includegraphics[scale=0.6]{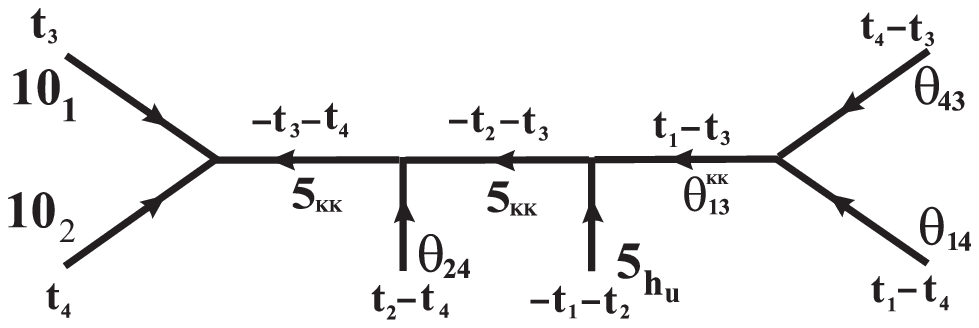}\;\;\;\;
\includegraphics[scale=0.6]{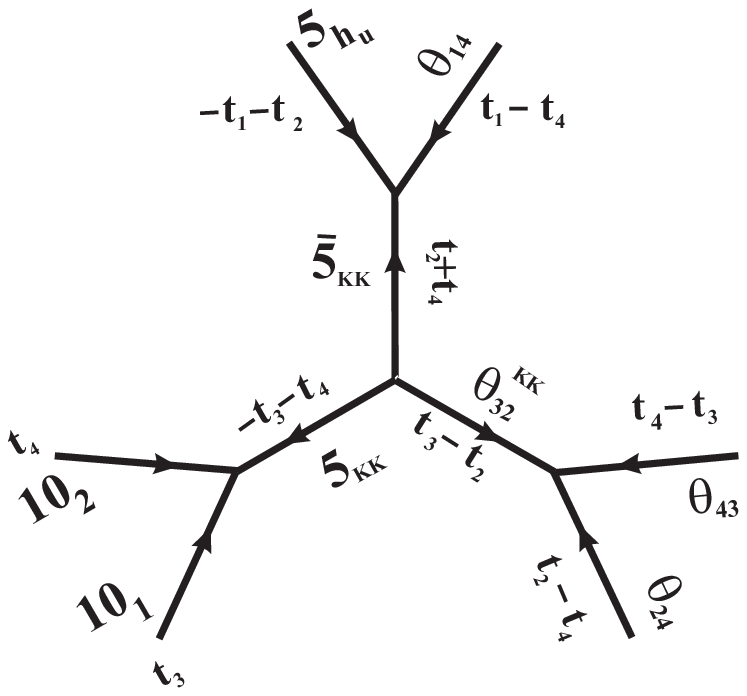}
\caption{Graphs generating the $\lambda^{u}_{12}$ up Yukawa coupling. Mass insertions in KK-mode lines are not shown to avoid clutter.}
 \label{up12}
\end{figure}

We have already calculated the tree-level coupling $\lambda_{33}^t\approx 0.31$. The remaining couplings are calculated in a manner similar to that for the down quark.   Examples of the relevant graphs are depicted in fig~\ref{up22} and \ref{up12}. The resulting up-quark mass matrix is
\ba
M_u&=&\left(
\begin{array}{lll}
0.09\,\theta_{14}^2\theta_{43}^2 & 0.22\,\theta_{14}^2\theta_{43} & 0.16\,\theta_{14}\theta_{43} \\
0.22\,\theta_{14}^2\theta_{43} & 0.18\,\theta_{14}^2 & 0.22 \,\theta_{14}\\
0.16\,\theta_{14}\theta_{43} & 0.22\,\theta_{14} & 0.31
\end{array}
\right)
\label{MUU}
\ea

\subsubsection{Further phenomenological considerations}

We found that in the case of the down quark mass matrix there are significant cancellations  in the determinant and subdeterminants that are independent of the normalisation of the fields.  The same is true in the up quark case.
 From eq.(\ref{MUU}) we have Det$(M_{u})=3\times 10^{-3}\theta_{14}^{4}\theta_{43}^{2}$ and Det$(M_{u}^{23})=8.10^{-2}\theta_{14}^{2}$ where the latter refers to the $2\times 2$ submatrix for the second and third generations that determines the product $m_{c}m_{t}$. For comparison if we impose the texture zeros in the $(1,1)$ and $(2,2)$ positions, that would follow if the matter curves determining these elements do not intersect, we find Det$(M_{u})=8\times 10^{-4}\theta_{14}^{4}\theta_{43}^{2}$ and Det$(M_{u}^{23})=4.6\times 10^{-2}\theta_{14}\theta_{43}$. These do not change as much as for the down quark mass matrix but the result is misleading because the $(1,1)$ matrix element is already very small, explaining the small difference between these cases. However the value of the determinant is small due to a cancellation of individual terms of $O(10^{-2})$, demonstrating again that the F-theory structure of the Yukawa couplings does lead to significant cancellations that can have important phenomenological effects.

Knowing the up and down quark mass matrices we can compute the masses and mixing angles in terms of the two familon expectation values. For definiteness we choose to have texture zeros in the (1,1) and (2,2) entries of both the up and down quark mass matrices (the general structure is similar if we do not have texture zeros in the up matrix). Keeping the same values for the familon vevs as were used in Section \ref{4.3.1}, we find $m_{c}/m_{t}=3.9\times 10^{-3}$, $m_{u}/m_{t}=10^{-4}$, $V_{cb}=10^{-3}$ and $V_{us}=0.12$. Of these only the value of $m_{c}$ agrees with the measured value at the Planck scale. The value of $m_{u}$ is a factor of 10 too large while the mixing angles are too small as a result of significant cancellations between the up and down sectors. However, due to the strong cancellation in Det$(M_{u})$, the associated up quark mass is very sensitive to small corrections. Indeed changing $M_{u}(3,3)$ from $0.31$ to $0.33$ brings $m_{c}$ into good agreement with experiment (corrections of this magnitude are expected from radiative threshold corrections \cite{Pierce:1996zz}). Similarly it is straightforward to bring the mixing angles into good agreement with experiment by spoiling the precise cancellation between the up and down mixing angles. Changing $M_{d}(3,2)$ from $0.17$ to $0.1$ and $M_{d}(2,3)$ from $0.21$ to $0.35$ gives $V_{cb}=0.042$ in excellent agreement with experiment. Changes of this magnitude are expected from the $\tan\beta$ enhanced threshold corrections \cite{Pierce:1996zz}. Similarly one may get an acceptable value for $V_{us}$ by changing $M_{d}(1,2)$ from $0.1$ to $0.2$. Since the radiative corrections to this element do not involve the large third generation Yukawa coupling they are very small so a change of this magnitude must come from another source; for example, {\it c.f.} figs \ref{down12} and \ref{up12}, if  the $\theta_{13}^{KK},\;\theta_{32}^{KK}$ messenger masses are a factor of 2 different from the other messenger masses.

In conclusion the structure of couplings found in F-theory raises interesting possibilities for generating realistic masses and mixings. In particular because the magnitude of the couplings in various elements of the mass matrix are correlated there can be significant cancellations between terms and this can play an important role in generating the hierarchical mass structure. This is to be contrasted from the ``bottom-up'' Abelian family symmetry models of fermion masses in which the $O(1)$ couplings are not determined and which assume significant cancellations do not occur. Similarly the mixing angles in the F-theory model involve significant cancellations between the up and the down sectors. For the particular model considered here we have shown that small corrections to the Yukawa couplings determined assuming a very symmetric structure for the compactification manifold can give an excellent fit to all the quark masses and mixing angles.

\subsubsection{Charged leptons}\label{chlep}

A major problem with the $SU(5)$ model discussed here is the difficulty in obtaining a lepton mass matrix of a different form from the down quark mass matrix. While the relation $m_{b}=m_{\tau}$ may be acceptable at the GUT scale \cite{Ross:2007az} the relations $m_{s}=m_{\mu}$ and $m_{d}=m_{e}$ are not. We are aware of two possible resolutions to the problem. The first is that the GUT group is broken by flux raising the possibility that the flux may split the $SU(5)$ representations so that down quarks and leptons may live on different matter curves. However this possibility was apparently excluded  by Dudas and Palti \cite{Dudas:2009hu} who extended the no-go theorem of ~\cite{Marsano:2009wr} to the case the families live on different matter curves. Recently they have suggested ways to evade this theorem \cite{Dudas:2010zb} at the same time solving the doublet-triplet splitting problem. In the next section we present a simpler example of this mechanism. We also discuss how the method can be applied to splitting the down quarks and leptons in the second generation only. In both cases the ``good'' relation $m_{b}=m_{\tau}$ is preserved while the ``bad'' relation $m_{\mu}=m_{s}$ can be avoided. However the remaining ``good'' relation $Det[M_{d}=Det[M_{l}]$ is also  not guaranteed by the underlying GUT symmetry.

The second possibility is that radiative threshold corrections spoil the ``bad'' mass relations while preserving both the ``good'' mass relations. That this is a possibility follows from two features of the F-theory mass matrices. The first is the near equality of the top and the bottom Yukawa couplings. This means that radiative corrections to the down quark Yukawa couplings involving the third generation will be large. In particular there are large threshold ``corrections'' \cite{Pierce:1996zz} coupling the bottom quarks to the up Higgs conjugate field via an intermediate $t^{c}$ state, resulting in a $\tan\beta$ enhanced contribution to the mass matrix. There is no such contribution to the lepton mass matrix because the singlet neutrino states, $\nu^{c}$ are heavy. The second feature is the possibility of cancellation in the sub-determinants of the Yukawa matrix that we discussed above. To illustrate this we return to the down quark mass matrix of eq(\ref{MDD}) which has the (2,3) subdeterminant Det$(M_{d}^{23})=10^{-2}\theta_{14}\theta_{43}$. Even a small threshold correction to the mass matrix can change this sub-determinant significantly. For example a reduction of $M_{d}^{33}$ by $15\%$ reduces Det$(M_{d}^{23})$ by a factor of 3 and a $\tan\beta$ enhanced correction of this size is to be expected. Such a reduction will imply that $m_{\mu}=3m_{s}$ at the unification scale in excellent agreement with the measured masses once the RG corrections are included. The  $15\%$ reduction in  Det$(M_{d}^{23})$ implies $m_{\tau}=1.15m_{b}$ so the GUT relation is not changed much and is in reasonable agreement with the measured values. However the change in $M_{d}^{33}$ does change Det$(M_{d})$ significantly spoiling a ``good'' relation. To avoid this we can require a texture zero in $M_{d}^{11}$ without affecting the $(2,3)$ sector. This avoids significant cancellation in Det$(M_{d})$ with the result that it is only reduced by $15\%$ by the $15\%$ reduction in $M_{d}^{33}$. This implies that $m_{e}=m_{d}/3$, again in excellent agreement with experiment.
This example illustrates the possibility of using threshold corrections to avoid the ``bad'' $SU(5)$ relation for the second generation quark and lepton masses. However it cannot be applied to the particular model studied here because it relies on the strong cancellation in a sub-determinant that in this particular model is inconsistent with the quark mass hierarchy;  the model above avoided the cancellation by imposing texture zeros.

\subsubsection{Neutrino masses}
We turn now to a discussion of the neutrino masses in this model. As pointed out in~\cite{Bouchard:2009bu}  in models with a monodromy there will be Standard Model singlet states with Majorana mass and these may generate Majorana mass for the doublet neutrinos via the see-saw mechanism. For the model introduced here the states $\theta_{ij}'$ have the quantum numbers to be right-handed neutrinos. Due to the monodromy $\theta_{12}'$ and $\theta_{21}'$ are identified so, in the covering theory, the superpotential term $M_{M}\theta_{12}'\theta_{21}'$ that is allowed by all the symmetries, is a Majorana mass in the quotient theory. This state couples to the doublet neutrinos via the terms $(L_{3}\theta_{14}+L_{2}\theta_{14}\theta_{43}+L_{1}\theta_{14}^{2}\theta_{43})H_{u}\theta_{12}'$ and, through the see-saw mechanism, will generate a Majorana mass for the combination
\[ L_{3}\langle\theta_{14}\rangle+L_{2}\langle\theta_{14}\rangle \langle\theta_{43}\rangle +L_{1}\langle\theta_{14}\rangle ^{2}\langle\theta_{43}\rangle .\]
 However this combination does not correspond to the near bi-maximal mixing observed for the (heaviest) atmospheric neutrino and moreover doesn't indicate how the solar neutrino acquires a mass.
\begin{figure}[h]
\centering
\includegraphics[scale=0.8]{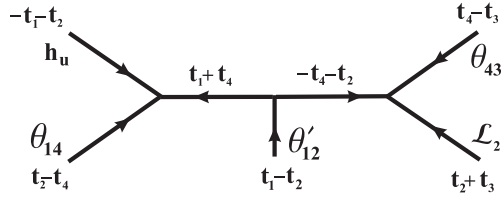}
\caption{A representative graph for the  neutrino Yukawa coupling.}
 \label{d13}
\end{figure}

To address these questions we consider the case that there are additional vectorlike pairs of RH neutrino states, $\theta_{14}',\theta_{41}'$ and $\theta_{13}',\theta_{31}'$ with mass $M_{D}$. These may be part of a tower of Kaluza Klein states or zero mode pairs that acquire a mass when $SU(5)$ is broken. In addition there is a mass mixing term given by $\theta_{14}'\theta_{31}'\langle\theta_{43}\rangle$. For $\langle\theta_{43}\rangle\ll M_{D}$ the mass eigenstates are $\theta_{a,b}'=(\theta_{41}'\pm\theta_{31}')/\sqrt{2}$ with mass $M_{a,b}^{2}=M_D^{2}\pm M_{D}\langle\theta_{43}\rangle$.
The states $\theta_{a,b}'$  acquire Majorana mass via their mixing with $\theta_{12}'$ through a see-saw mechanism involving the couplings $(\theta_{41}'\langle\theta_{14}\rangle+\theta_{31}'\langle\theta_{43}\theta_{14}\rangle)\theta_{12}'$. This in turn generates a Majorana mass for a combination of the light LH neutrinos. However if two combinations of LH neutrino states are to acquire  Majorana mass it is necessary that there should be two primary Majorana mass terms and these states should couple to different combinations of the LH neutrino states. Provided there is more than one $\theta_{12}'$ KK state on the matter curve the first condition will be satisfied. Because the KK wavefunctions are not holomorphic  the coupling of the LH neutrinos to these KK states will be different~\cite{Bouchard:2009bu}, but comparable, satisfying the second condition. As a result both $\theta_{a}$ and $\theta_{b}$ acquire Majorana masses of $O(\langle\theta_{14}\rangle^2/M_{M})$. For the case that the Majorana mass $M_{M}$ is large this term is smaller than $M_{D}$.

The coupling of $\theta_{a,b}$ to the light neutrinos is via the terms \[\left(L_{3}\theta_{14}'+L_{2}\theta_{13}'+L_{1}(\langle\theta_{43}\theta_{14}\rangle\theta_{14}'+\langle\theta_{14}\rangle\theta_{13}'\right)\,H_{u}\]
The LH neutrinos now acquire Majorana masses via the see-saw mechanism by coupling to  the messenger states $\theta_{a,b}'$ and for small $M_{D}$ these can be dominant.
\subsection{Bi-maximal mixing}
The atmospheric neutrino state is the heaviest and its mass will be generated by the exchange of the lightest RH neutrino, $\theta_{a}'$, so
 the atmospheric state is
\ba
\nu_{@}\approx \frac{1}{\sqrt{2}}\left[\nu_{3}-\nu_{2}+\sqrt{2}\langle\theta_{14}\rangle(\langle\theta_{43}\rangle-1))\nu_{1}\right ]
\ea
with mass
\ba
m_{@}=\frac{\langle h_{u}\rangle^{2}}{M_D^{2}-M_{D}\langle\theta_{43}\rangle}\frac{\langle\theta_{14}\rangle^{2}}{\sqrt{2}M_{M}}.
\ea
As one may see the state is nearly maximally mixed between $\nu_{\mu}$ and $\nu_{\tau}$ in good agreement with experiment.
\subsection{Tri-maximal mixing}
The exchange of the next lightest RH neutrino state generates $\nu_{\odot}$, the next heaviest LH neutrino. Here it is the state $\theta_{b}'$ giving
\ba
\nu_{\odot}\approx \frac{1}{N}\left[\nu_{3}+\nu_{2}+\sqrt{2}\langle\theta_{14}\rangle(1+\langle\theta_{43}\rangle)\nu_{1}\right]
\ea
with mass
\ba
m_{@}=\frac{\langle h_{u}\rangle^{2}}{M_D^{2}+M_{D}\langle\theta_{43}\rangle}\frac{\sqrt{3}\langle\theta_{14}\rangle^{2}}{{2}M_{M}}.
\ea
The mixing between all three neutrinos is large but not clearly tri-maximal with the $\nu_{e}$ component somewhat smaller than the experimental measurement. However it is sensitive to the mass of the messenger states responsible for the higher dimension terms ($\langle\theta_{14,43}\rangle\equiv \langle\theta_{14,43}\rangle/M_{messenger}$). It only requires a small difference in messenger mass between the quark mass sector and the neutrino see-saw sector to allow for tri-maximal mixing here .

\subsubsection{CP violation}\label{CP}

So far we have considered the case that the Yukawa couplings and the associated mass matrices are real. However it is important to identify the possible origin of the CP violating phase. An obvious possibility is that the familon fields have complex vevs. As may be seen from the Appendix their phases are not determined by the D-terms. They  can be induced by the soft supersymmetry breaking terms. However it is straightforward to check that, in the case of the model considered above with just two familon fields contributing to the mass matrices, the resultant phases appearing in eqs \ref{massmatrices} can be eliminated by a phase redefinition of the fermion fields. This is no longer true when the threshold effects are included. As discussed above these can be large because the model is at large $\tan \beta$ but even so it is unlikely that the CP violation is near maximal as is observed. It is possible that the relative magnitude of the CP violation coming from threshold effects is enhanced in the case there are significant cancellations between the different terms in the mass matrix but this is not the case for the model considered because of the texture zeros that were assumed to avoid such cancellations.

There are two other possible sources of CP violation.
The first is due to flux effects. In the next Section we consider the case that flux causes the Higgs doublet triplet splitting. In this case two of the $SU(2)$ singlet up quarks and charge leptons belong to a single up quark and charged lepton matter curve respectively.  As a result some of the Yukawa couplings vanish in the absence of flux. When flux is switched on these couplings can acquire phases that cannot be removed by field redefinition and thus generate CP violation. These phases can be of $O(1)$ even though the flux effects are small because they are associated with couplings whose magnitudes also vanish in the absence of flux.

The second possibility follows from the fact, noted in Section \ref{ZM}, that the solution to the 8D equs involves an arbitrary holomorphic function $f(z_{L})$ of the longtitudinal coordinate along the matter curve.  For the case that the fields live on different matter curves and the hierarchy is organised by the family symmetry there is no need for the intersections of the matter curves  to occur at the same point. This is to be contrasted with the case that the three families live on the same matter curve where the intersections should be close together to avoid large mixing angles due to the mismatch of the up and down sectors. Given this, the value of $f(z_{L})$ associated with the intersection point of a given up quark Yukawa coupling  can be different from the value associated with the intersection point associated with the equivalent down quark Yukawa coupling. The difference between these factors can introduce a phase difference between these couplings that cannot be absorbed by a redefinition of the fermion fields and can be the source of CP violation. Of course it can also introduce a difference in the magnitude of the Yukawa couplings and this will change the estimate of the Yukawa coupling presented above.

\section{Fluxes - doublet triplet splitting and distinguishing between down quarks and leptons}\label{flux}

In this section we explore the use of fluxes to derive different textures of the down and lepton mass matrices and to split doublets from triplets in the Higgs sector. This is based on the fact that fluxes may split the $SU(5)$ representations and break the quark-lepton degeneracy in the Kaluza Klein states. An example of this was given in \cite{Dudas:2010zb} in which there are a large number of vector-like exotic states that have to be given mass by familon vevs.

Here we give an example with the minimum number of vector-like states capable of generating doublet-triplet splitting. It necessarily requires that the right-handed electron be in a different representation from the down quark, thus allowing them to have different masses. In this example the difference between the muon and strange quark masses must be due to the non-degeneracy of the KK states. The only GUT relation that persists in this case is that relating the bottom quark to the tau lepton.

We also give an example in which the second family is split so that the muon and strange quark belong to different representations and as a result they need not be degenerate. In this case the GUT relation for the the ratio of the electron to the down quark mass  is preserved up to the $O(1)$ couplings. We determine this ratio for the very symmetric compactification discussed above.

\begin{table}[tbp] \centering%
\begin{tabular}{|l|c|c|c|c|}
\hline
Field&$U(1)_i$& homology& $U(1)_Y$-flux&$U(1)$-flux\\
\hline
$10^{(1)}=10_3$& $t_{1,2}$& $\eta-2c_1-{\chi}$&$ -N$ &$M_{10_1}$\\ \hline
$10^{(2)}=10_1$& $t_{3}$& $-c_1+\chi_7$&$ N_7$ &$M_{10_2}$\\ \hline
$10^{(3)}=10_2$& $t_{4}$& $-c_1+\chi_8$&$ N_8$ &$M_{10_3}$\\ \hline
$10^{(4)}=10_2'$& $t_{5}$& $-c_1+\chi_9$&$ N_9$ &$M_{10_4}$\\ \hline
$5^{(0)}=5_{h_u}$& $-t_{1}-t_2$& $-c_1+{\chi}$&$ N$ &$M_{5_{h_u}}$\\ \hline
$5^{(1)}=5_2$& $-t_{1,2}-t_3$& $\eta -2c_1-{\chi}$&$ -N$ &$M_{5_1}$\\ \hline
$5^{(2)}=5_3$& $-t_{1,2}-t_4$& $\eta -2c_1-{\chi}$&$ -N$ &$M_{5_2}$\\ \hline
$5^{(3)}=5_x$& $-t_{1,2}-t_5$& $\eta -2c_1-{\chi}$&$ -N$ &$M_{5_3}$\\ \hline
$5^{(4)}=5_1$& $-t_{3}-t_4$& $-c_1+{\chi}-\chi_9$&$N-N_9$ &$M_{5_4}$\\ \hline
$5^{(5)}=5_{h_d}$& $-t_{3}-t_5$& $-c_1+{\chi}-\chi_8$&$ N-N_8$ &$M_{5_{h_d}}$\\ \hline
$5^{(6)}=5_y$& $-t_{4}-t_5$& $-c_1+{\chi}-\chi_7$&$ N-N_7$ &$M_{5_6}$\\ \hline
\end{tabular}%
\caption{Field representation content under $SU(5)\times U(1)_{t_i}$, their homology
class and flux restrictions. For convenience, only the properties of $10,5$ are shown.
$\ov{10},\ov{5}$ are characterized by opposite values of $t_i\ra -t_i$ etc.
Note that the fluxes satisfy $N=N_7 +  N_8 + N_9$ and $\sum_iM_{10_i}+\sum_jM_{5_j}=0$
while  ${\chi}=\chi_7 +  \chi_8 + \chi_9$.}
\label{rest}
\end{table}

The starting point for model building is given in Table~\ref{rest} adapted from \cite{Dudas:2010zb}. The first two columns give the field content under $SU(5)\times U(1)_{t_i}$ for the case of $Z_{2}$ monodromy.  The third column gives the Dudas Palti determination of the homology classes where $c_{1}$ is the first Chern class of the tangent bundle of GUT surface $S_{GUT}$ and $\eta=6c_{1}-t$ with $-t$ being the first Chern class of the normal bundle to $S_{GUT}$. The $\chi_{i}$ are unspecified and ${\chi}=\chi_7 +  \chi_8 + \chi_9$.

When $U(1)_{Y}$ hypercharge flux is turned on the $SU(5)$ multiplets can be split to Standard Model multiplets
\ba
10&\ra& (3,2)_{\frac 16}+(\bar 3,1)_{-\frac 23}+(1,1)_1\nn\\
5&\ra& (3,1)_{-\frac 13}+(1,2)_{\frac 12}
\ea

Given a restriction to a curve of the  $U(1)_{Y}$ hypercharge flux given by an integer $N_{Y}$ and the $U(1)_{t_i}$ flux  given by an integer $M_{10,5}$ the resulting spectrum is
\ba
n_{(3,2)_{\frac 16}}-n_{(\bar 3,2)_{-\frac 16}}&=&M_{10}\\
n_{(\bar 3,1)_{-\frac 23}}-n_{(3,1)_{\frac 23}}&=&M_{10}-N_Y\label{10dec}\\
n_{(1,1)_{1}}-n_{(1,1)_{-1}}&=&M_{10}+N_Y
\ea
and
\ba
n_{(3,1)_{-\frac 13}}-n_{(\bar 3,1)_{\frac 13}}&=&M_5\\
n_{(1,2)_{\frac 12}}-n_{(1, 2)_{-\frac 12}}&=&M_{5}+N_Y\label{5dec}
\ea
Depending on the particular flux properties, different $10$'s and $5$'s have
also different $M_{10},M_{5}$. Due to anomaly cancelation these are constrained by \ba
\sum_iM_{10_i}+\sum_jM_{5_j}&=&0
\ea

\subsection{A simple model for doublet-triplet splitting}

To obtain the three complete families of Table \ref{Reps} we need $M_{10_{1,2,3}}=1$, $M_{5_{1,2,4}}=-1$ and $N=0$. Turning to the Higgs structure we need $M_{5_{0}}=1$.  It is possible to have a split $\bar{5}_{h_{u}}$ with only the doublet massless mode by choosing $M_{5_{5}}=0$ and $N_8=1$ giving

\[
n_{(3,1)_{ - 1/3} }  - n_{(\overline 3 ,1)_{1/3} }  = M_{5_5 }  = 0
\]
\[
n_{(1,2)_{1/2} }  - n_{(1,2)_{ - 1/2} }  = M_{5_5 }  + N - N_8  =  - 1.
\]

Next we satisfy the trace conditions by choosing $M_{5_6 }  =  - 1,\;N_7  =  - 1$
so that $\bar{5}_{6}$ has only a colour triplet component:
\[
n_{(3,1)_{ - 1/3} }  - n_{(\overline 3 ,1)_{1/3} }  = M_{5_6 }  =  - 1
\]
\[
n_{(1,2)_{1/2} }  - n_{(1,2)_{ - 1/2} }  = M_{5_6 }  + N - N_7  =  0.
\]
To complete the doublet triplet splitting we can give a mass $m_{T}=<\theta_{15}><\theta_{14}>/M$ to the anti-triplet in $5_{h_{u}}$ by coupling it to this triplet via the superpotential term $\theta _{15} \theta _{14} 5_{h_u } \overline 5 _y/M $. Now the only light Higgs fields are the doublets in $5_{h_{u}}$ and $\bar{5}_{h_{d}}$ as desired.

However from Table \ref{Reps} one may see that the $10_{2,3}$ representations will be affected by the $N_{7,8}$ flux. The $10_{2}$ representation is split according to
\[
n_{(3,2)_{1/6} }  - n_{(\overline 3 ,2)_{ - 1/6} }  = M_{10_2 }  = 1
\]
\[
n_{(\overline 3 ,1)_{ - 2/3} }  - n_{(3,1)_{2/3} }  = M_{10_2 }  - N_7  = 2
\]
\[
n_{(1,1)_1 }  - n_{(1,1)_{ - 1} }  = M_{10_2 }  + N_7  = 0
\]
and the $10_{3}$ representation is split according to
\[
n_{(3,2)_{1/6} }  - n_{(\overline 3 ,2)_{ - 1/6} }  = M_{10_3 }  = 1
\]
\[
n_{(\overline 3 ,1)_{ - 2/3} }  - n_{(3,1)_{2/3} }  = M_{10_3 }  - N_8  = 0
\]
\[
n_{(1,1)_1 }  - n_{(1,1)_{ - 1} }  = M_{10_3 }  + N_8  = 2.
\]
The result of this is that the up quark and lepton mass matrices change and now have the form
\[
M_u  = \left( {\begin{array}{*{20}c}
   {\varepsilon _1 \theta _{14} ^2 \theta _{43}^2 } & {\varepsilon _2 \theta _{14} ^2 \theta _{43}^{} } & {\varepsilon _3 \theta _{14} ^{} \theta _{43}^{} }  \\
   {\theta _{14} ^2 \theta _{43}^2 } & {\theta _{14} ^2 \theta _{43}^{} } & {\theta _{14} ^{} \theta _{43}^{} }  \\
   {\theta _{14} ^{} \theta _{43}^{} } & {\theta _{14} } & 1  \\
\end{array}} \right)
\]
and
\[
M_l  = \left( {\begin{array}{*{20}c}
   {\varepsilon _1 '\theta _{14} ^2 \theta _{43}^{} } & {\varepsilon _2 '\theta _{14} ^{} \theta _{43}^{} } & {\varepsilon _3 '\theta _{14} }  \\
   {\theta _{14} ^2 \theta _{43}^{} } & {\theta _{14} ^{} \theta _{43}^{} } & {\theta _{14} }  \\
   {\theta _{14} ^{} \theta _{43}^{} } & {\theta _{43} } & 1  \\
\end{array}} \right).
\]
where  $\epsilon_{i}, \epsilon'_{i}$  are (small) coefficients generated by flux because now 2 up quark and 2 lepton fields live on the same quark and lepton curves respectively so only one field has a Yukawa coupling to a given state in the absence of flux.

The change in  $M_{u}$  enhances the hierarchy in the up quark mass matrix as it is the smaller row that is duplicated. As a result it should be easier to fit the observed spectrum without requiring cancellations in the determinant and sub-determinants. The change in  $M_{l}$ affects $m_{e}$  so it can be different from $m_{d}$ but not, to first order, $m_{\mu}$. However the effect of the flux on the matter curve $5^{(5)}=5_{h_{d}}$ not only splits the zero mode multiplet but also removes the KK excitation degeneracy. As a result the higher dimension operators in the $(2,3)$ block of $M_{d}$ and $M_{l}$ will have different messenger masses and so $m_{\mu}$ can be different from $m_{s}$. In this case only the $(3,3)$ element is unaffected by flux because, coming from a renormalisable term, only it is insensitive to the KK mass spectrum.  This means that the GUT relation $m_{b}=m_{\tau}$ is preserved, up to the threshold corrections discussed in Section \ref{chlep}.

There are two potential problems associated with this method of doublet triplet splitting. The first is the fact that residual heavy colour triplet state of mass $m_{T}$ affects the gauge coupling unification. However, as shown in~\cite{Blumenhagen:2008aw,Marsano:2009wr,Leontaris:2009wi} the effect of these states is to compensate for the flux correction to the usual $SU(5)$ predictions so that for a particular choice of the colour triplet mass (dependent on the flux strength) the MSSM form of gauge coupling unification is obtained. If the colour triplet mass is less than this the corrections act to reduce the residual discrepancy between the predicted and observed strong coupling thus actually improving the agreement over the MSSM~\cite{Pokorski:1997mv}.

The second potential problem arises from the need for a large vev for the familon field $\theta_{15}$ to generate a mass for the colour triplet component in $5_{h_{u}}$. One must check that it does not introduce D=5 nucleon decay operators and this in fact is the case due to $t_{5}$ conservation because $\theta_{15}$ only has a negative $t_{5}$ component. This is in contrast to the example given in \cite{Dudas:2010zb}. As noted in Section \ref{mu} a vev for $\theta_{15}$  generates a $\mu$ term via the operator $\lambda H_uH_d\theta _{14}\theta_{43}\theta_{15}$. Provided the fields involved in generating this term do not intersect the coefficient $\lambda$ can be small and the term can provide the needed origin for the $\mu$ term.

\subsection{Splitting the second family}

Our second example has $M_{10_{1,2,3}}=1$, $M_{5_{1,2,4}}=-1$ as before. In addition $N_9=1=-N_8$. These choices are consistent with the restrictions discussed above. Concentrating first on
the two $10$'s that are affected by these non-trivial fluxes, we get the following solutions
for the multiplicities of the various $10,5$ components:
\ba
10^{(3)}({t_4})=10_2:&\left\{\begin{array}{lll}
         n_{(3,2)_{\frac 16}}-n_{(\bar 3,2)_{-\frac 16}}=1&\ra & 1\times Q\\
          n_{(\bar 3,1)_{-\frac 23}}-n_{(\bar 3,1)_{\frac 23}}=2&\ra& 2\times u^c\\
         n_{(1,1)_{1}}-n_{(1,1)_{-1}}=0&\ra& 0\times e^c
         \end{array}\right.\label{t4comp}
         \\
 10^{(4)}({t_5})=10_2':&\left\{\begin{array}{lll}
         n_{(3,2)_{\frac 16}}-n_{(\bar 3,2)_{-\frac 16}}=0&\ra & 0\times Q\\
          n_{(\bar 3,1)_{-\frac 23}}-n_{(\bar 3,1)_{\frac 23}}=-1&\ra& 1\times \bar u^c\\
         n_{(1,1)_{1}}-n_{(1,1)_{-1}}=1&\ra& 1\times e^c
         \end{array}\right.\label{t5comp}
         \ea
We observe that eqs.(\ref{t4comp},\ref{t5comp}) split the
second generation into two distinct tenplets, however the cost to pay is the presence
of an additional $u^c,\bar{u}^c$ pair. We can give mass to this extra-pair through the
coupling
\[ {\cal W}_{nr}\supset \frac{1}{M}\,\ov{10}'_2\cdot 10_2\cdot\langle\theta_{51}\theta_{14}\rangle\;=\;\frac{M_{\theta}^2}{M}\,\bar{u}^c\,u^c\]

An analogous splitting occurs in the fiveplets $\bar{5}^{(4)}=\bar 5_1$ and $\bar{5}^{(5)}=\bar 5_{h_d}$.
Choosing $M_{5_4}=-1,M_{5_{h_d}}=0$ and, given the values $N_{8,9}$ above, we get
\ba
5^{(4)}({-t_3-t_4})=5_1:&\left\{\begin{array}{lll}
          n_{(\bar 3,1)_{-\frac 23}}-n_{(\bar 3,1)_{\frac 23}}=-1&\ra& 1\times d^c\in \bar 5^{(4)}\\
         n_{(1,1)_{1}}-n_{(1,1)_{-1}}=-2&\ra& 2\times \ell\in \bar 5^{(4)}
         \end{array}\right.\label{5t4comp}
         \\
 5^{(5)}({-t_3-t_5})=5_{h_d}:&\left\{\begin{array}{lll}
          n_{(\bar 3,1)_{-\frac 13}}-n_{(\bar 3,1)_{\frac 13}}=0&\ra& 0\times \bar D_h^c\\
         n_{(1,2)_{\frac 12}}-n_{(1,2)_{-\frac 12}}=1&\ra& 1\times \bar\ell\in 5^{(5)}
         \end{array}\right.\label{5t5comp}
         \ea
We can interpret this as follows: the curve accommodating the down-Higgs fiveplet has
an excess  of an anti-doublet  $\bar\ell$ which can form one massive state with one
of the two $\ell$'s of $\bar 5^{(4)}$-curve
\[ {\cal W}_{nr}\supset \frac{1}{M}\,\ov{5}^{(4)}\cdot 5^{(5)}\cdot\langle\theta_{51}\theta_{14}\rangle\;=\;\frac{M_{\theta}^2}{M}\,\bar{\ell}\,\ell\]
The remaining massless chiral state $\ell\in 5_1$ is part of curve $5^{(4)}$
accommodating the first family and, together with the down quark $d^c\in 5_1$
constitute a complete anti-fiveplet.
\bigskip

This splitting of the second family  differentiates the lepton mass matrix relative to down quark mass matrix.
The substitution $10_{t_4}\ra 10_{t_5}$ for the RH lepton of second generation  ($\mu^c$)
gives a lepton mass matrix of the form
\ba
M_{\ell}&=&\left(
\begin{array}{lll}
\theta _{14}^2 \theta _{43}^2 &  \theta_{15}\theta_{14}\theta_{43} &  \theta _{14} \,\theta _{43} \\
 \theta _{14}^2 \theta _{43} & \theta_{15}\theta_{43} &\theta _{14} \\
\theta _{14} \theta _{43} &\theta _{15} & 0.29
\end{array}
\right)
\ea
Comparing with the form of the down quark mass matrix given in eq.(\ref{MDD}) one sees that, as in the last example, the  GUT relation $m_{b}=m_{\tau}$ is preserved. However the muon mass is proportional to $\theta_{15}\theta_{43}$ whereas the strange quark is proportional to $\theta_{14}\theta_{43}$ so the masses at the GUT scale need not obey the (bad) GUT mass relation.

This example does not achieve doublet-triplet splitting by flux and requires that it be achieved through discrete Wilson line breaking of $SU(5)$. If this is possible it has the advantage of not having additional massive colour triplet states that can spoil gauge coupling unification. However there is a problem is associated with the dimension 5 nucleon decay operators. These have the structure $10.10.10.\bar{5}$ and are forbidden before the familon field $\theta_{51}$ acquires a vev because, {\it c.f.} Table \ref{Reps}  there is no $10$ or $\bar{5}$ field involving $t_{5}$. However operators of the form  $\theta_{51}Q_{3}Q_{3}Q_2L_{2}^{c}/M^{2}$ are allowed and potentially cause the proton decay rate to exceed the observed bound.  As discussed in \cite{Dudas:2010zb} these operators do not arise if the strong version of R-conservation is applied because they are not generated by KK mediator graphs.

\section{Summary}
Although the calculation of fermion masses and mixing remains an elusive goal, F-theory offers some new insights because the trilinear couplings in the superpotential associated with the intersection of three matter fields depend on local details only and thus can be reliable calculated even if the global structure of the theory is not known.
In this paper we have derived the general form for the trilinear couplings involving zero modes alone or involving both zero modes and massive modes. The resulting Yukawa couplings are found to be simple analytic functions of the matter field Abelian charges and  the GUT-value of the unified gauge coupling constant raised to  a fractional power.

To illustrate the implications of such couplings we considered a particular model with a $SU(5)\otimes SU(5)_{\perp}$ gauge symmetry and three families living on different matter curves with a $Z_{2}$ monodromy to allow for an hierarchy of fermion masses and a $Z_{2}$ R-symmetry to avoid rapid proton decay. The gauge symmetry of the resulting model is $SU(5)\otimes U(1)^{3}$ where the $U(1)$ factors play the role of family symmetries. The hierarchy of fermion masses and mixings is then generated through an expansion in the familon vevs breaking the symmetry. The masses and mixing of the lighter families occur through operators involving the familon fields with dimension greater than four. The calculation of the coefficients of these operators requires knowledge of the massive messenger sector and in calculating these we assumed a very symmetric structure for the representations of these fields and for the global structure of the compactification manifold.

The resulting structure shows some interesting features not normally considered in family symmetry models. In particular the determinants and subdeterminants of the mass matrices are anomalously small due to a cancellation between terms indicating correlation between the calculated $O(1)$ coefficients of the mass matrix elements. The mixing angles are also subject to large cancellations between the up and down quark contributions. In the absence of a theory of the $O(1)$ couplings this possibility is usually ignored as,  it requires ``fine tuning''. In F-theory models such cancellations do occur and opens up new possibilities for model building.

In the case of the trial model the structure of the model was chosen assuming no such cancellations and to avoid them two texture zeros were assumed corresponding to the non-intersection of the relevant matter fields. Taking account of threshold corrections and varying the two relevant familon vevs the model allows for an excellent fit of all quark masses and mixing angles with the exception of the up quark mass. Fitting the latter requires a modest difference in the messenger masses associated with the operators generating the up quark mass matrix elements.

In theories with a monodromy Majorana masses are generated for singlet states with the quantum numbers of right-handed neutrinos~\cite{Bouchard:2009bu}. In the trial model we showed that, via the see-saw mechanism,  these generate light neutrino masses with small mass differences caused by flux effects. Near tri-bi-maximal mixing readily arises providing a structure in good agreement with that found in neutrino oscillation.

Finally we considered the question how to modify the $SU(5)$ ``bad'' relations connecting down quark masses and leptons. For the case that there are significant cancellations in the down quark mass matrix determinants, threshold corrections can generate the difference between the muon mass and the strange quark mass while preserving the ``good'' $SU(5)$ mass relations for the first and the third families.  For the case of the trial model considered here  this explanation is not possible and one must rely on flux effects to do the job. We derived a simple example of a flux choice that  generates doublet-triplet splitting
 leaving only the required Higgs doublets in the light spectrum. The effect on the massive sector then spoils the GUT predictions for the light generations that obtain their mass from non-renormalisable operators sensitive to the structure of the massive sector. We also constructed an example in which the second generation is split so that the second generation down quarks and leptons belong to different matter curves. Again this preserves the good GUT mass relation for the heavy family but corrects both the good and the bad relations  for the first and second families.

 In conclusion we have considered the detailed spontaneously broken family symmetry structure emerging from F theory for the case that the three families live on separate matter curves. Assuming that the compactification and flux structure leads to R-symmetry and choosing an appropriate multiplet structure we  showed that the resulting model can fit all the observed quark, charged lepton and neutrino mass and mixing angle structure. It remains to be seen whether a global version exists with this structure.

 \bigskip

 \noindent{\bf{\Large Acknowledgement}}

 We would like to thank Emilian Dudas and Eran Palti for useful discussions.
This work is partially supported by the European Research and Training Network (RTN)
grant ``Unification in the LHC era ''(PITN-GA-2009-237920).

 \bigskip

\noindent{\bf{\Large Appendix : The familon potential in $SU(5)\times U(1)^3$}}

 In order to generate the $\mu$ term and the fermion masses discussed above it is necessary to generate vevs for the familon fields $\theta_{14},\theta_{15}$ and $\theta_{43}$.  In fact it is the expectation that three familon fields should acquire vevs because, after imposing the monodromy,  there are three $U(1)$ gauge bosons and satisfying  D-flatness for them typically requires three charged fields to acquire vevs. To see this, consider the basis in the covering theory given in eq(\ref{Qbasis}). The $U(1)$ gauge boson associated with $Q_{2}$ is projected out by the monodromy and the D-flatness conditions for the remaining three $U(1)$ gauge bosons are

 \ba
\left|\langle\theta_{14}\rangle\right|^2+\left|\langle\theta_{15}\rangle\right|^2+\xi_1&=&0\nn\\
-\left|\langle\theta_{43}\rangle\right|^2-\left|\langle\theta_{15}\rangle\right|^2+\xi_3&=&0\nn\\
-2\left|\langle\theta_{14}\rangle\right|^2-3\left|\langle\theta_{43}\rangle\right|^2+\left|\langle\theta_{15}\rangle\right|^2+\xi_4&=&0\nn
\label{DT}
\ea
where the $\xi_{i}$ are the anomalous contributions to the $U(1)$s in this basis.  For $\xi_{1}$ negative, $\xi_{3}$ positive and $\xi_{4}<3\xi_{3}-2\xi_{1}$ there is a $D$-flat direction along which the three familon fields acquire vevs. Provided there is no $\theta_{31}$ zero mode this set of vevs is also F-flat.

 \end{document}